\documentclass[aps,prb,reprint,superscriptaddress]{revtex4-2}
\usepackage{amssymb}
\usepackage{amsfonts}
\usepackage{xcolor}

\newcommand{\orcid}[1]{\href{https://orcid.org/#1}{\textsuperscript{[ORCID]}}}
\usepackage{amsmath}
\usepackage{comment}
\usepackage{braket}
\usepackage{booktabs}
\usepackage{graphicx}

\providecommand{\Tr}{\operatorname{Tr}}

\providecommand{\Eprint}[1]{#1}

\providecommand{\Eq}[1]{Eq.~(\ref{#1})}

\begin{document}

\title{Information Hierarchy in Many-Body Berry Phase}%

\author{Kai Watanabe}
\email{kaiw.sshhrm@gmail.com} 
 \affiliation{Independent Researcher} 
\date{\today}

\begin{abstract}
Many-body topology is a central concept in modern theories of solids, and identifying effective degrees of freedom that capture it is important both fundamentally and practically.
This work studies the extent to which geometric information of an interacting many-body ground state can be inferred from a finite number of local correlations.
Starting from the Resta formula, $ z=\left\langle \exp\!\left(\frac{2\pi i}{L}\hat X\right)\right\rangle$, we view $\log z$ as the cumulant generating function and establish a generic information hierarchy across cumulant orders.
We show that, for an $N$-particle system, even complete knowledge of all density correlators up to order $N-1$ does not, in general, uniquely determine the Berry phase $\gamma=\operatorname{Im}\log z \, (\mathrm{mod}\ 2\pi)$.
In the thermodynamic limit, the statement becomes stronger: no finite set of local correlators suffices to determine the global holonomy.
We also identify two exceptional yes-go cases in which the hierarchy is broken.
First, for quasi-free models, all cumulants are determined by the particle two-point correlation function.
Second, symmetry-enforced constraints can reduce the infinite cumulant sum entering $\log z$ to finite information.
The argument is analytic and does not rely on a specific microscopic Hamiltonian.
Our results clarify a limitation of approaches based on local degrees of freedom for many-body holonomy and provide a minimal framework for distinguishing when global holonomies are encoded in local correlations and when they are not.
We also comment on the possibility of analogous hierarchies in other contexts,
such as the quantum marginal problem in quantum information theory and
many-body scattering problems.
Finally, we discuss implications for future numerical work, including
machine-learning approaches to the search for topological phases.
\end{abstract}

\maketitle
\section{Introduction}
In modern theories of solids, topological structures of quantum many-body systems determine their responses to gauge fields and are related to global observables such as polarization~\cite{NiuThoulessWu1985,Vanderbilt2018BerryPhasesBook,Berry1984}.
Thus, describing such holonomies in terms of effective degrees of freedom is important both practically and fundamentally.
Practically, it provides a basis for developing economical yet reliable computational techniques for many-body systems based on essential set of degrees of freedom~\cite{DiSante2026KagomeMetalsRMP,Schmidt_ET,Sommer_Wen_Vishwanath_HBC_I_2025PRL}.
Fundamentally, this description paves the way to identify the relevant degrees of freedom underlying topological structures, from crystals to polarization phenomena of the quantum vacuum.
The importance of finding the degrees of freedom determining the holonomy is not confined to condensed matter physics but extends to other fields that deal with quantum many-body systems, for example, nuclear theory and quantum field theory~\cite{tHooft1978,Luscher1986ScatteringStates,Wilson1974}.

A representative holonomy in condensed matter physics is the Berry phase.
Using the Resta formula~\cite{Resta1998PositionOperatorPBC}, the Berry phase is
given by the argument of the expectation value of the Lieb-Schultz-Mattis (LSM)
unitary operator constructed from density operators.
Thus, one may naively expect that it can be computed once density correlators up
to some finite order are known.
When symmetries are present or when a one-body approximation is valid, either
exactly or approximately, the Berry phase can indeed be related to reduced
density information~\cite{RequistGross2018Polarization,GiesbertzRuggenthaler2019PhysRep,GoriGiorgiZiesche2002PRB}.
Such quantities have also been widely used to characterize topological
structures~\cite{Chiu2016RMP}.

However, a closer look at prior works suggests that the Berry phase may not be
determined by local bulk correlators alone.
In a periodic ring, the total $U(1)$ flux threading the ring cannot be gauged away.
Nevertheless, one can choose a gauge in which the flux is concentrated near the
boundary, namely, a twisted boundary condition.
In this gauge, local bulk observables in the gapped regime can become
insensitive to the flux twist angle in the thermodynamic
limit~\cite{WatanabeHA}.
On the other hand, the Berry phase is the global phase acquired by the ground
state when the flux twist angle $\theta$ is varied from $0$ to $2\pi$.
This raises the possibility that local measurements in the bulk do not contain
sufficient information to determine the Berry phase.
This point is closely related to the observation by Ortiz, Souza, and
Martin~\cite{OrtizSouzaMartin1998ExchangeCorrelationHole} that the Berry-phase
polarization cannot be inferred solely from local bulk density information.

Another indication comes from the LSM unitary~\cite{Bohm1949Bloch,LiebSchultzMattis1961} appearing in the Resta formula.
A Taylor expansion of this unitary generates statistical moments of the multipole operator, whose local definition under periodic boundary conditions is known to be subtle~\cite{OnoTrifunovicWatanabe2019}.
This may suggest that truncating the expansion at a finite order hides a conceptual issue.

A closely related obstruction appears in quantum information theory.
A many-body state can always be reduced to subsystem states, for example through partial traces or a Schmidt decomposition across a bipartition.
However, reconstructing the original many-body state from such reduced states requires additional global consistency conditions.
This issue is known as the quantum marginal problem~\cite{Yu2021NatCommun}.
In many-body physics, it is related to the $N$-representability problem, which concerns the reconstruction of an $N$-body quantum state from reduced density matrices~\cite{Liu2007PRL}.

These considerations do not by themselves prove that the Berry phase cannot be determined from local degrees of freedom.
However, they suggest that there are intrinsic limitations to local information.
How should one reconcile the many existing examples where the Berry phase is extracted from local correlators~\cite{GiesbertzRuggenthaler2019PhysRep,SokolovSchaefer2013ODCFT}?
This work addresses this question by asking to what extent the Berry phase can be constrained by local correlation functions.

More concretely, we regard the logarithm of the Resta formula as a cumulant generating function~\cite{SouzaWilkensMartin2000,RestaSorella1999,HetenyiCengiz2022}.
Then, we address the above question by examining cumulants order by order.
In this formulation, the $p$-th cumulant is determined by connected density correlators up to $p$ points.
If the Berry phase is fixed at some finite order $p=r$, then it is determined from density correlators of finite order.
On the other hand, if higher-order cumulants are not controlled, it suggests that the Berry phase cannot be determined from any finite set of correlators.

The paper is organized as follows.
In Sec.~\ref{sec:Nogo}, we interpret the Resta formula for the Berry phase as a cumulant generating function and derive the cumulants order by order.
We first present a simple example that illustrates the order-by-order information hierarchy as a warm-up.
We then establish a generic hierarchy in a model-independent manner. This yields
a no-go statement that, for an $N$-particle system, density correlators up to
order $N-1$ do not determine the Berry phase in general.
In the thermodynamic limit, this statement is elevated to a local-to-global
hierarchy: no fixed finite order of local density correlators determines the
Berry phase in general.
In Sec.~\ref{sec:yesgo}, we identify two yes-go scenarios.
The first is the quasi-free case, namely ground states of quadratic Hamiltonians.
The second is symmetry-induced information reduction, which breaks the information hierarchy.
In Sec.~\ref{sec:discussion}, we discuss connections to prior works, including the quantum marginal problem in quantum information theory.
We also explain why previous approaches based on density-correlator calculations
can succeed in symmetry-enforced yes-go scenarios.
We also discuss guidelines for machine-learning approaches to topological phases.
Then, we briefly comment on the possibility that scattering phase shifts in interacting many-body systems may be viewed as global phases and may exhibit analogous information hierarchies.
Finally, we summarize our results and outline future directions in Sec.~\ref{sec:summary}.

\section{The information hierarchy from the cumulant expansion}
\label{sec:Nogo}
\subsection{Resta formula as a cumulant generating function}
We start from the Resta formula~\cite{Resta1998PositionOperatorPBC}, given by
\begin{equation}
z=\langle \Psi|\hat U_\mathrm{LSM}|\Psi\rangle.
\label{eq:z_def}
\end{equation}
Here, the unitary $\hat U_\mathrm{LSM} = e^{i\alpha_L \hat X}$ is  the Lieb-Schultz-Matis (LSM) unitary~\cite{Bohm1949Bloch,LiebSchultzMattis1961,YamanakaOshikawaAffleck1997,Oshikawa2000} and $\hat X=\sum_{j=1}^{L} j\,\hat n_j$ is the generator of the local $U(1)$ transformation with the angle $\alpha_L=\frac{2\pi}{L}$ \footnote{To be accurate, the local charge $\hat n_j$ is the generator of a local $U(1)$ gauge rotation. Accordingly, $\hat X=\sum_{j=1}^{L} j\,\hat n_j$ is a linear combination of these generators and generates a position-dependent phase rotation.}.
In $\hat X$,  $\hat n_j$ denotes the density operator at site $j$.
We omit spin indices because they are not relevant in the present discussion.
The density operator is defined in the conventional way in terms of the electron annihilation and creation operators,
$\hat c_j$ and $\hat c_j^\dagger$, as
\[
\hat n_j=\hat c_j^\dagger \hat c_j .
\]
The vacuum state $|0\rangle$ is defined by
\[
\hat c_j|0\rangle=0
\qquad
\text{for all } j .
\]
The fermion operators satisfy the canonical anticommutation relations
\[
\{\hat c_i,\hat c_j^\dagger\}=\delta_{ij},
\qquad
\{\hat c_i,\hat c_j\}=0,
\qquad
\{\hat c_i^\dagger,\hat c_j^\dagger\}=0 .
\]
Accordingly,
\[
[\hat n_j,\hat c_j]=-\hat c_j,
\qquad
[\hat n_j,\hat c_j^\dagger]=\hat c_j^\dagger .
\]
The many-body state $|\Psi\rangle$ is an element of the Fock space generated by these fermion creation operators from the vacuum.

The formula in Eq.~\eqref{eq:z_def} is thus the expectation value of the LSM unitary evaluated in the many-body state $|\Psi\rangle$.
When $|\Psi\rangle$ is a gapped and nondegenerate many-body ground state, the argument of $z$ can be interpreted as the Berry phase.
Thus,
\begin{equation}
\gamma=\operatorname{Im}\log z \quad (\mathrm{mod}\ 2\pi).
\label{eq:gamma_def}
\end{equation}
The Resta formula ~\Eq{eq:z_def} can be viewed as a shortcut to the Wilson-loop evaluation of the Abelian Berry phase associated with the $U(1)$ flux-insertion cycle $\theta:0\to 2\pi$ ~\cite{Berry1984,watanabe2026exactconjugationidentitymanybody}.
The total flux $\theta$ threading the ring cannot be gauged out by any single-valued
$U(1)$ transformation under the periodic boundary condition (PBC).
It is therefore a gauge-invariant parameter of the Hamiltonian.
The phase accumulated along a closed $\theta$-cycle is the Berry phase.

We assume that the ground state remains gapped and nondegenerate throughout the $\theta$-cycle so that the Berry phase is well-defined.
The proof that such gapped ground states exist is given in Appendix~\ref{app:gap_and_seam}, where we employ the technique of H.~Watanabe~\cite{WatanabeHA,WatanabeHB}.

The key observation is that the logarithm of $z$ in Eq.~\eqref{eq:z_def} is a cumulant generating function in the sense of mathematical statistics:
\begin{equation}
\log z=\sum_{p\ge 1}\frac{(i\alpha_L)^p}{p!}\kappa_p.
\label{eq:logz_cumulant}
\end{equation}
The $p$-th order cumulant of $\hat X$ in the state $|\Psi\rangle$ is given by
\begin{equation}
\kappa_p=\left.\frac{d^p}{dt^p}\log\langle \Psi|e^{t\hat X}|\Psi\rangle\right|_{t=0}.
\label{eq:kappa_def}
\end{equation}
Here and below, we do not put a hat on $\kappa_p$, since it is a c-number.
For instance, the first three cumulants are given by
\begin{equation}
\kappa_1=\langle \Psi|\hat X|\Psi\rangle.
\label{eq:kappa1}
\end{equation}
\begin{equation}
\kappa_2=\langle \Psi|\hat X^2|\Psi\rangle-\kappa_1^2.
\label{eq:kappa2}
\end{equation}
\begin{equation}
\kappa_3=\langle \Psi|\hat X^3|\Psi\rangle-3\kappa_2\kappa_1-\kappa_1^3.
\label{eq:kappa3}
\end{equation}
Since $\hat X$ is a weighted sum of commuting density operators, the $p$-th cumulant $\kappa_p$ is controlled by $p$-point connected density correlators, namely, $\langle \Psi|\hat n_{j_1}\hat n_{j_2}\cdots \hat n_{j_p}|\Psi\rangle$.
In what follows, we denote the expectation value with respect to $|\Psi\rangle$ by $\langle\cdot\rangle$, namely,
$\langle \hat O\rangle=\langle \Psi|\hat O|\Psi\rangle$,
to save space, if not mentioned.

Now, we show that fixing cumulants up to order $p$ does not, in general, constrain cumulants of order $p+1$ and higher.
We refer to this as the cumulant order-by-order hierarchy non-closure.
Since Eq.~\eqref{eq:logz_cumulant} extends to all orders, this implies that finite number of density correlators (equivalently, finitely many cumulants) do not determine $\gamma$ in general.
Equivalently, the many-body Berry phase can not be, in general, expressed as a functional of any fixed finite local correlations.

\subsection{Minimal example from $L=6$ chain ring at half filling}
We begin with an analytically tractable example, for which all quantities can be evaluated explicitly and the cumulant order-by-order hierarchy non-closure mechanism is visualized.

For $L=6$, the ground state can be expanded as a linear combination of the occupation basis, namely,
\begin{equation}
|\Psi\rangle=\sum_{n_1,\dots,n_6\in\{0,1\}} a_{n_1\cdots n_6}\,|n_1 n_2 \cdots n_6\rangle.
\label{eq:Psi_occ_expansion_L6}
\end{equation}
At half filling $N=\sum_{j=1}^{6}<\hat n_j>=3$, there are 20 basis states.
Among them, we consider two pure states as uniform superpositions of four $N=3$ configurations:
\begin{equation}
|\Psi_E\rangle=\frac12\Bigl(|000111\rangle+|011100\rangle+|101010\rangle+|110001\rangle\Bigr).
\label{eq:PsiE_L6}
\end{equation}
\begin{equation}
|\Psi_O\rangle=\frac12\Bigl(|001110\rangle+|010101\rangle+|100011\rangle+|111000\rangle\Bigr).
\label{eq:PsiO_L6}
\end{equation}

The two states of interest are chosen so that the one- and two-point density correlators obey particularly simple analytic constraints.
For both states, all density one-point correlators coincide:
\begin{widetext}
\begin{equation}
\langle \Psi_E|\hat n_i|\Psi_E\rangle=\langle \Psi_O|\hat n_i|\Psi_O\rangle=\frac12
\qquad (\forall\, i\in\{1,\dots,6\}).
\label{eq:L6_1pt}
\end{equation}
\end{widetext}
Moreover, all density two-point correlators read
\begin{widetext}
\begin{equation}
\langle \Psi_E|\hat n_i\hat n_{i+3}|\Psi_E\rangle=\langle \Psi_O|\hat n_i\hat n_{i+3}|\Psi_O\rangle=0
\qquad (\forall\, i\in\{1,2,3\}).
\label{eq:L6_2pt_opposite}
\end{equation}
\begin{equation}
\langle \Psi_E|\hat n_i\hat n_j|\Psi_E\rangle=\langle \Psi_O|\hat n_i\hat n_j|\Psi_O\rangle=\frac14
\qquad (\forall\, i,j\in\{1,\dots,6\},\ i\neq j,\ j\neq i+3).
\label{eq:L6_2pt_other}
\end{equation}
\end{widetext}

The cumulants up to second order are evaluated from the density correlators.
For example, the first-order cumulant reads
\begin{equation}
\kappa_1=\langle \hat X\rangle=\sum_{j=1}^{6} j\,\langle \hat n_j\rangle.
\end{equation}
Thus, substituting Eq.~\eqref{eq:L6_1pt}, we arrive at
\begin{equation}
\kappa_1\big|_{\Psi_E}=\kappa_1\big|_{\Psi_O}=\frac{21}{2}.
\label{eq:L6_kappa1_value}
\end{equation}
The calculation for the second-order cumulant is also straightforward.
The result is
\begin{equation}
\kappa_2\big|_{\Psi_E}=\kappa_2\big|_{\Psi_O}=\frac{27}{4}.
\label{eq:L6_kappa2_value}
\end{equation}
Thus, the cumulants from the even and odd ground states coincide up to second order.

However, the third-order cumulant is not identical and has opposite signs.
This is because the density three-point correlator for the even and odd states differ.
Explicitly, for the even ground state, the three-point correlator is given by,
\begin{widetext}
\begin{equation}
\bigl\langle \Psi_E |\hat n_{j_1}\hat n_{j_2}\hat n_{j_3}|\Psi_E\bigr\rangle
=
\begin{cases}
\displaystyle \frac{1}{4}
\qquad \mathrm{for}\ (j_1,j_2,j_3)\in\{(4,5,6),(2,3,4),(1,3,5),(1,2,6)\},\\[4pt]
0,  \qquad \text{otherwise}.
\end{cases}
\end{equation}
\end{widetext}
In the odd ground state, the site configurations for which the three-point density correlator takes a finite value differ from those in the even case, which reads
\begin{widetext}
\begin{equation}
\bigl\langle \Psi_O |\hat n_{j_1}\hat n_{j_2}\hat n_{j_3}|\Psi_O \bigr\rangle
=
\begin{cases}
\displaystyle \frac{1}{4} 
\qquad \mathrm{for}\  (j_1,j_2,j_3)\in\{(3,4,5),(2,4,6),(1,5,6),(1,2,3)\},\\[4pt]
0,  \qquad \text{otherwise}.
\end{cases}
\end{equation}
\end{widetext}
Substituting the above together with the first and second order cumulants, we find,
\begin{equation}
\kappa_3\big|_{\Psi_E}=\frac{81}{4},
\label{eq:L6_kappa3_value_E}
\end{equation}
and
\begin{equation}
\kappa_3\big|_{\Psi_O}=-\frac{81}{4}.
\label{eq:L6_kappa3_value_O}
\end{equation}
Thus, the constraints on density one-point and two-point correlators do not constrain the third order cumulant.

The three-point correlators differ due to the hidden particle-number parity.
To see this, we partition the ring into two three-site blocks, $B_1=\{1,2,3\}$ and $B_2=\{4,5,6\}$.
The two states in Eqs.~\eqref{eq:PsiE_L6} and \eqref{eq:PsiO_L6} are distinguished by the number parity on $B_1$:
$|\Psi_E\rangle$ and $|\Psi_O\rangle$ yield $(-1)^{n_1+n_2+n_3}=+1$ and $-1$, respectively.
The three-point correlator encodes this number parity and is therefore not constrained by one-point and two-point correlators.
This limitation of low-order correlators is consistent with the discussion of the one-particle reduced-density-matrix case~\cite{SchillingSchilling2016}.

From the above, the Berry phases computed from the even and odd ground states start to differ at third order in the cumulant expansion.
The difference between the Berry phase evaluated with the even ground state (denoted by $\gamma_E$) and that evaluated with the odd ground state (denoted by $\gamma_O$) reads
\begin{equation}
\gamma_E-\gamma_O
=-\frac{27}{4}\, \alpha_L^3+O(\alpha_L^5).
\label{eq:L6_gamma_diff}
\end{equation}
Here, the $O(\alpha_L^5)$ terms arise from $\kappa_5$ and higher cumulants.
Terms of even order do not contribute since they are real; moreover, for this example they vanish exactly, as shown in the discussion below.

Higher order cumulants  $O(\alpha_L^5)$ are not constrained either.
To see this, we evaluate the Resta formula in terms of cumulants of $\hat X$ in each state, one may write
\begin{equation}
\log z_E=\sum_{p\ge 1}\frac{(i\alpha_L)^p}{p!}\kappa_p\big|_{\Psi_E}.
\label{eq:logzE_cumulant}
\end{equation}
\begin{equation}
\log z_O=\sum_{p\ge 1}\frac{(i\alpha_L)^p}{p!}\kappa_p\big|_{\Psi_O}.
\label{eq:logzO_cumulant}
\end{equation}
Therefore,
\begin{equation}
\log\frac{z_E}{z_O}
=\sum_{p\ge 1}\frac{(i\alpha_L)^p}{p!}\Bigl(\kappa_p\big|_{\Psi_E}-\kappa_p\big|_{\Psi_O}\Bigr).
\label{eq:log_ratio_cumulant}
\end{equation}

In the present $L=6$ example, $\hat X$ takes only two values on the occupation configurations appearing in each state.
For $|\Psi_E\rangle$, one has $X\in\{9,15\}$, and hence
\begin{equation}
X_E=9+6Y,
\label{eq:X_Bernoulli_E}
\end{equation}
where $Y\in\{0,1\}$ is a Bernoulli random variable with $\mathbf{P}(Y=1)=1/4$.
For $|\Psi_O\rangle$, one has $X\in\{6,12\}$.
Using the same Bernoulli variable $Y$, this can be written as
\begin{equation}
X_O=12-6Y .
\label{eq:X_Bernoulli_O}
\end{equation}
Thus, the even and odd cases are related by the sign reversal of the $Y$ term.

Writing the random variables explicitly, for $p\ge 2$, their cumulants satisfy
\begin{equation}
\begin{cases}
\displaystyle
\kappa_p[6Y+9]
=
6^p\kappa_p[Y],
\\[0.6em]
\displaystyle
\kappa_p[-6Y+12]
=
(-6)^p\kappa_p[Y].
\end{cases}
\label{eq:cumulant_affine_property_EO}
\end{equation}
Though a constant shift affects only the first cumulant, i.e., the average, in the present construction, the first cumulants are already fixed to be identical.
As a result, the odd-order cumulants change sign while the even-order cumulants coincide,
which implies that $\log(z_E/z_O)$ receives contributions only from the odd-order cumulants as
\begin{equation}
\log\frac{z_E}{z_O}
=2\sum_{m\ge 1}\frac{(i\alpha_L)^{2m+1}}{(2m+1)!}\,\kappa_{2m+1}\big|_{\Psi_E}.
\label{eq:L6_logz_ratio_odd}
\end{equation}

One might naively expect cancellations after summing over all orders in the cumulant expansion.
However, $z_E$ and $z_O$ can be evaluated directly from the occupation configurations.
For $\alpha_L=\pi/3$, this yields $z_E=-1$ and $z_O=+1$, and equivalently, $\gamma_E=\pi$ and $\gamma_O=0$, respectively.
This shows that the difference between the two states persists to all orders and is not removed by any potential cancelation.

Although all density correlators on four distinct sites vanish in the $N=3$ sector, this does not imply the vanishing of higher cumulants.
Indeed, expanding $\hat X^p$ generates many terms with repeated indices, and the identity $\hat n_i^2=\hat n_i$ reduces such terms to lower-point correlators, which is finite.

\subsection{General case with arbitrary $L$ and $N$}
Now we shall move on to the general case with arbitrary $L$ and $N$.
The ground state in the occupation basis is given by
\begin{equation}
|\Psi\rangle=\sum_{n_1,\dots,n_L\in\{0,1\}} a_{n_1\cdots n_L}\,|n_1\cdots n_L\rangle,
\label{eq:general_ground_states}
\end{equation}
the coefficient $a_{n_1\cdots n_L}\,|n_1\cdots n_L\rangle$ satisfy $\sum_{n_1,\dots,n_L}|a_{n_1\cdots n_L}|^2=1$ and thus $p(n_1,\dots,n_L):=|a_{n_1\cdots n_L}|^2$ is the weight.
We abbreviate $\mathbf n:=(n_1,\dots,n_L)\in\{0,1\}^L$ and write $p(\mathbf n)$.
To fix the particle number sector, the weight is set to $p(\mathbf n)=0$ unless $\sum_{j=1}^L n_j=N$.

Since $\hat n_j$ and hence $\hat X=\sum_{j=1}^L j\hat n_j$ are diagonal in this basis, we obtain
\begin{equation}
z:=\langle \Psi|e^{i\alpha_L \hat X}|\Psi\rangle
=\sum_{\mathbf n} p(\mathbf n)\,e^{i\alpha_L X(\mathbf n)},
\end{equation}
where $X(\mathbf n):=\sum_{j=1}^L j\,n_j$ is the expectation value of $\hat X$. 
The density $p$-point correlator is given by
\begin{equation}
\langle \hat n_{j_1}\cdots \hat n_{j_p}\rangle
=\sum_{\mathbf n} p(\mathbf n)\,n_{j_1}\cdots n_{j_p},
\qquad (1\le p\le r) .
\end{equation}
Constraining the $p$-point correlation function is equivalent to fixing the expectation values, that is,
\begin{equation}
\langle \hat n_{j_1}\cdots \hat n_{j_p}\rangle
=A_{j_1\cdots j_p},
\label{eq:n-point_constrain}
\end{equation}
for all choices of $p$ sites $(j_1,\dots,j_p)$, where $A_{j_1\cdots j_p}$ is a constant.

We now emphasize that fixing density correlators up to order $p$ does not uniquely specify the ground state.
Indeed, there exist infinitely many superposition states in the occupation basis whose weights are chosen so that
they share the same set of correlators $A_{j_1\cdots j_p}$ for all $p\le r$ (and for all choices of sites).

The above is explicitly shown by the cumulant expansion.
Let $|\Psi\rangle$ and $|\Psi'\rangle$ be two gapped ground states in the same $(L,N)$ sector.
Then, we have two expectation values of the LSM unitary given by,
\[
z=\langle\Psi|\hat U_{\mathrm{LSM}}|\Psi\rangle,\qquad
z'=\langle\Psi'|\hat U_{\mathrm{LSM}}|\Psi'\rangle .
\]
Assume that density correlators coincide up to order $r$,
\begin{equation}
\label{eq:moment_match}
\langle\Psi|\hat n_{j_1}\cdots \hat n_{j_p}|\Psi\rangle
=
\langle\Psi'|\hat n_{j_1}\cdots \hat n_{j_p}|\Psi'\rangle,
\qquad (1\le p\le r),
\end{equation}
for all choices of sites $(j_1,\dots,j_p)$.

Under the condition~\eqref{eq:moment_match}, the cumulant expansion yields
\begin{equation}
\label{eq:cumulant_difference_general}
\log\frac{z}{z'}
=
\sum_{p=r+1}^{\infty}\frac{(i\alpha_L)^p}{p!}\,\bigl(\kappa_p-\kappa_p'\bigr),
\end{equation}
where $\kappa_p$ and $\kappa_p'$ are the $p$-th cumulants of $\hat X$ evaluated in $|\Psi\rangle$ and $|\Psi'\rangle$, respectively.
Since cumulants with $p\ge r+1$ are not determined by correlators up to order $r$, the right-hand side of
\eqref{eq:cumulant_difference_general} is unconstrained in general.
Note that the cancellation of even orders observed in the special case $(L,N)=(6,3)$ relies on additional structure where a hidden particle-number parity leads to a Bernoulli statistics.

\subsubsection{Scaling with system size $L$}
One may naively expect that higher-order cumulants are suppressed since $\alpha_L=2\pi/L$ is small for large system size $L$.
However, cumulants of $\hat X$ naturally involve the combination $\alpha_L j$, which is typically
\[
\alpha_L j \sim O(1)\qquad (j\sim O(L)).
\]
Thus the scaling from $\alpha_L$ is compensated by the scale of $j$, so that each order can remain $O(1)$.
Moreover, the $p$-th order contribution involves a sum over choices of $p$ sites, whose combinatorial factor scales as
\[
\binom{L}{p}\sim O(L^{p}).
\]
Therefore, no suppression in $L$ is guaranteed, and the cumulant expansion is not controlled solely by the smallness of $\alpha_L$.

Table~\ref{tab:cumulant_check} summarizes the exact cumulants for the
equal-weighted superposition of all configurations in the half filling sector
for representative system sizes $L$.
Here, the cumulants were evaluated from the central moments~\cite{HetenyiCengiz2022}
by exhaustively enumerating all configurations in the half filling sector.
The table shows that the higher-order cumulants grow rather than being suppressed.
Consequently, the factor $(i\alpha_L)^p/p!$ does not make the rescaled cumulants
negligible, as seen from the fourth- and sixth-order values.
Moreover, the cumulants themselves grow with increasing $L$.

In the thermodynamic limit $L\to\infty$, the particle number $N$ also diverges, and the cumulant hierarchy is naturally promoted from an order-by-order hierarchy to a local-to-global hierarchy.
At a fixed order $p$, the number of density correlators already scales as
$\binom{L}{p}\sim O(L^p)$.
Thus, even before taking the order $p$ to increase with $N$, the number of
density correlators that must be fixed in order to determine the Berry phase
grows polynomially with the system size.

As shown in the next section, however, this obstruction can be avoided in at
least two exceptional scenarios, where additional structures make a practical
determination possible.

An alternative proof of the hierarchy is given in
Appendix~\ref{app:factorization}, where we write $z$ explicitly in terms of
density $p$-point correlators rather than cumulants.

\begin{table}[t]
\centering
\caption{
Exact cumulants for the equal-weighted superposition of all configurations in the half-filling sector.
}
\label{tab:cumulant_check}
\begin{ruledtabular}
\begin{tabular}{cc@{\qquad}|ccc}
$L$ & $N$ & $p$ & $\kappa_p$ &
$\textstyle \frac{(i\alpha_L)^p}{p!}\kappa_p$ \\
\hline
 &  & $2$ &
$\textstyle \frac{21}{4}$ &
$\simeq -2.8786$
\\
$6$ & $3$ & $4$ &
$\textstyle -\frac{693}{40}$ &
$ \simeq -0.8681$
\\
& & $6$ &
$\textstyle \frac{1041}{4}$ &
$\simeq -0.4767$
\\
\hline
 &  & $2$ &
$12$ &
$ \simeq -3.7011$
\\
$8$ & $4$ & $4$ &
$\textstyle -\frac{336}{5}$ &
$ \simeq -1.0654$
\\
& & $6$ &
$\textstyle \frac{12144}{7}$ &
$ \simeq -0.5655$
\\
\hline
 &  & $2$ &
$39$ &
$\simeq -5.3460$
\\
$12$ & $6$ & $4$ &
$-468$ &
$ \simeq -1.4656$
\\
& & $6$ &
$\textstyle \frac{183378}{7}$ &
$\simeq -0.7497$
\\
\hline
 &  & $2$ &
$\textstyle \frac{272}{3}$ &
$ \simeq -6.9910$
\\
$16$ & $8$ & $4$ &
$\textstyle -\frac{28288}{15}$ &
$ \simeq -1.8687$
\\
& & $6$ &
$\textstyle \frac{1655936}{9}$ &
$ \simeq -0.9372$
\end{tabular}
\end{ruledtabular}
\end{table}
\clearpage

\section{Exceptional cases where the hierarchy is reduced}
\label{sec:yesgo}
We have shown that, in general, the many-body Berry phase cannot be determined from density correlators up to a fixed order $p$.
However, as is also clear from numerous existing studies, there are many situations in which the Berry phase is successfully extracted from a finite set of density correlators \cite{KingSmithVanderbilt1993,GiesbertzRuggenthaler2019PhysRep,GoriGiorgiZiesche2002PRB}.

The origin of this apparent tension can be traced back to two distinct mechanisms, which we refer to as the yes-go scenario.
The first scenario is when Wick's theorem applies, so that higher-order
cumulants are fully determined by the particle two-point correlation function.
In the alternative case, the symmetry reduces the information hierarchy drastically.

\subsection{Gaussian closure}
The first exceptional case is that of quasi-free ground states.
The Hamiltonian is then quadratic in fermions and reads
\begin{equation}
  \hat H=\sum_{i,j= 1}^{L}\hat c_i^{\dagger}\,h_{ij}\,\hat c_j.
  \label{eq:quasi-free_hamiltonian}
\end{equation}
Here, $h_{ij}$ is a Hermitian one-body matrix and contains no density-dependent terms.
When the Hamiltonian is given in the form \Eq{eq:quasi-free_hamiltonian}, the Berry phase is determined from particle (not density) two-point correlators.

To see this, it is convenient to work in the finite-temperature formulation.
First, the unitary in the Resta formula can also be written in a quadratic form as
\begin{equation}
e^{i\alpha_L\hat X}
=\exp\!\left(i\alpha_L\sum_{i,j=1}^{L}\hat c_i^{\dagger}\,M_{ij}\,\hat c_j\right),
\end{equation}
where $M=\mathrm{diag}(1,2,\dots,L)$.
The expectation value of the LSM unitary is given by
\begin{equation}
z(\beta)=\frac{\Tr\!\left(e^{-\beta \hat H}\,e^{i\alpha_L \hat X}\right)}{\Tr\!\left(e^{-\beta \hat H}\right)},
\end{equation}
where $\beta$ is the inverse temperature.
Since both $\hat H$ and $\hat X$ are quadratic in fermions, the product
$e^{-\beta \hat H}e^{i\alpha_L \hat X}$ is again Gaussian and can be written as
\[
e^{-\beta \hat H}e^{i\alpha_L \hat X}
=
C\,e^{\left(
-\sum_{i,j}\hat c_i^\dagger K_{ij}\hat c_j
\right)},
\]
with a constant $C$ and a one-body matrix $K_{ij}$.
Therefore, for a quasi-free Hamiltonian, the expectation value of LSM unitary $z(\beta)$ is expressed in terms of the single-particle correlation matrix as
\begin{equation}
  z(\beta)=\det\!\left(\,I_{ij}-C_{ij}(\beta)+\sum_{k=1}^{L}C_{ik}(\beta)\,\bigl(e^{i\alpha_L M}\bigr)_{kj}\right),
  \label{eq:z_in_thermal_2pt_func}
\end{equation}
where $C_{ij}(\beta)=\langle \hat c_i^\dagger \hat c_j\rangle_\beta$ is the thermal particle two-point correlator.

In gapped models, the ground state becomes dominant in the zero-temperature limit, similar to the ground state saturation in imaginary time approaches.
This is easily shown by the spectral decomposition
\begin{equation}
\hat H=\sum_{n\ge 0}E_n\,|\psi_n\rangle\langle \psi_n|,
\end{equation}
where $|\psi_n\rangle$ denotes the $n$-th excited state.
We set, without loss of generality, $E_0=0$ for the ground state $|\Psi\rangle$ and define the gap by $E_1>0$.
Then, for the LSM unitary, we can re-write as,
\begin{widetext}
\begin{equation}
\langle \hat U_{\mathrm{LSM}}\rangle_{\beta}
=\frac{\sum_{n\ge 0}e^{-\beta E_n}\langle \psi_n|\hat U_{\mathrm{LSM}}|\psi_n\rangle}{\sum_{n\ge 0}e^{-\beta E_n}}
=\frac{\langle \Psi|\hat U_{\mathrm{LSM}}|\Psi\rangle+e^{-\beta E_1}\langle \psi_1|\hat U_{\mathrm{LSM}}|\psi_1\rangle+\cdots}
{1+e^{-\beta E_1}+\cdots}.
\end{equation}
\end{widetext}
Taking the limit $\beta\to\infty$, the expectation value reduces to that of the ground state.
Thus, we recover the Resta formula in the zero temperature as $\lim_{\beta\to \infty} z(\beta)\to z$.
In what follows, we assume that the limit is already taken and we omit $\beta$ from the arguments.

The determinant formula in \Eq{eq:z_in_thermal_2pt_func} already implies that
the Berry phase is determined solely by the two-point correlators.
The connection to the cumulant hierarchy is as follows.
Take a density $p$-point correlator, for example.
In a quasi-free state, Wick's theorem implies that it is completely determined by the particle two-point correlator as
\begin{widetext}
\begin{equation}
\label{eq:density_wick_perm_plain}
\langle \hat n_{j_1}\hat n_{j_2}\cdots \hat n_{j_p}\rangle
=
\sum_{\mathrm{all\ permutations}}(-1)^{\lambda}\,
\prod_{m=1}^{p}\bigl\langle \hat c_{j_m}^\dagger \hat c_{j_{m'}}\bigr\rangle,
\qquad (j_1,\dots,j_p\ \text{all distinct}),
\end{equation}
\end{widetext}
where the permutation maps $(j_1,\dots,j_p)\mapsto (j_{1'},\dots,j_{p'})$, and $\lambda$ is the number of pairwise interchanges needed to realize the permutation.

The equation (\ref{eq:density_wick_perm_plain}) shows that any density correlators are evaluated solely from $C_{ij}$.
Consequently, the cumulant hierarchy is governed by $C_{ij}$ as well, which explains the closure in the Gaussian case and hence the unique determination of the Berry phase.

\subsection{Symmetry enforced information reduction}
Symmetry provides the other yes-go mechanism, because it constrains the many-body system itself.
Since symmetry imposes constraints across all orders of cumulants (equivalently, on density correlators of all orders), it can dramatically reduce the information hierarchy.

Let us demonstrate this mechanism for inversion symmetry.
From \Eq{eq:general_ground_states}, inversion symmetry can be imposed by demanding that the weights in the occupation basis are invariant under the mapping $\mathcal I:j\mapsto L+1-j$, namely,
\begin{equation}
p(\mathbf n)=p(\mathcal I\mathbf n).
\end{equation}
Moreover, one easily finds that the mapping of $\hat X$ is given by
\begin{equation}
X(\mathcal I\mathbf n)=(L+1)N-X(\mathbf n).
\end{equation}
Therefore, the expectation value $z$ satisfies
\begin{align}
z
&=\sum_{\mathbf n}p(\mathbf n)\,e^{i\alpha_L X(\mathbf n)}
=\sum_{\mathbf n}p(\mathcal I\mathbf n)\,e^{i\alpha_L X(\mathcal I\mathbf n)}\nonumber\\
&=\sum_{\mathbf n}p(\mathbf n)\,e^{i\alpha_L\bigl((L+1)N-X(\mathbf n)\bigr)} \nonumber\\
&=e^{i\alpha_L(L+1)N}\sum_{\mathbf n}p(\mathbf n)\,e^{-i\alpha_L X(\mathbf n)}\nonumber\\
&=e^{i\alpha_L(L+1)N}\,z^\ast.
\label{eq:inv_constraint_z}
\end{align}
Thus, $z$ satisfies a conjugation constraint, which strongly constrains the
cumulants at all orders.
Taking out the argument of z, it shows that the Berry phase $\gamma=\mathrm{Im}\,\log z\ (\mathrm{mod}\ 2\pi)$ satisfies
\begin{equation}
\gamma=\frac{\pi N}{L}
\quad \text{or}\quad
\gamma=\frac{\pi N}{L}+\pi
\qquad (\mathrm{mod}\ 2\pi).
\end{equation}
This shows that, when inversion symmetry is present, the Berry phase is determined from $L$ and $N$.

To make the constraint explicit, we shift the origin and introduce the centered
position operator
\[
\hat X_c=\sum_{j=1}^{L}\left(j-\frac{L+1}{2}\right)\hat n_j
=\hat X-\frac{L+1}{2}\hat N.
\]
This shift of the origin to $(L+1)/2$ absorbs the phase factor $e^{-i\alpha_L\frac{L+1}{2}N}$.
Then, the constraint \Eq{eq:inv_constraint_z} reduces to $z_c=z_c^\ast$, i.e., $z_c$ is real.

The above discussion implies that symmetry reduces the infinite cumulant sum entering $\log z$ into finite information.
In this sense, the information hierarchy is broken by symmetry.
From an information-theoretic perspective, quantization of a holonomy, e.g.,
the $\mathbb{Z}_2$ quantization~\cite{FuKane2006,KaneMele2005Z2}, can be viewed
as an extreme form of information compression, where a priori continuous data
are reduced to a discrete set of quantized possibilities, namely topological
invariants~\cite{TKNN1982}.

In the $(L,N)=(6,3)$ analytic example, the underlying Bernoulli distribution leads to a cancelation of the even order cumulants in $\log z$ and $z$ is pure imaginary.
Once inversion symmetry is imposed, however, $z$ is constrained to be real.
These two constraints are incompatible in this analytic setting, and the expectation value can collapse to $z=0$~\footnote{For an inversion-symmetric Hamiltonian with a nondegenerate gapped ground state, the ground state can be chosen to have a definite inversion parity.
In the present analytic construction, inversion pairs the even and odd states, so the symmetric combination $|\Psi_+\rangle\propto|\Psi_E\rangle+|\Psi_O\rangle$ is the natural ground state in the presence of inversion symmetry.
Evaluating the Resta expectation value in $|\Psi_+\rangle$ immediately gives $z_{\Psi_+}=z_E+z_O=0$ by cancellation.}, so the Berry phase extracted from $z$ becomes ill-defined.
This can be understood as a consequence of imposing two independent constraints on the information hierarchy, which overconstrain the problem and leave no admissible solution.

\section{Discussion}
\label{sec:discussion}
Viewed through the Resta formula, the Berry phase $\gamma=\mathrm{Im}\,\log z$ plays the role of a cumulant generating functional, and the cumulant expansion naturally defines an order-by-order information hierarchy.
This viewpoint is closely related to the quantum marginal problem in quantum information, which asks whether prescribed reduced density matrices are compatible with a global pure state~\cite{Yu2021NatCommun}.
It has been pointed out that reconstructing a many-body state from reduced local data faces a computational obstruction~\cite{Liu2007PRL}.
Our result indicates that an analogous information hierarchy exists in the context of topological many-body systems.
The appearance of related hierarchies in different fields may indicate a simple general principle:
although passing from the original many-body state to reduced subsystem state is well defined,
the inverse operation is not well defined without additional global consistency conditions.

However, physical many-body systems often have symmetries that  can break the information hierarchy as we have seen previously.
This is consistent with the symmetry-based organization of topological phases.
In many classification frameworks, global invariants are constrained and often reduced to discrete labels once symmetries are imposed \cite{ChiuRMP2016,ShiozakiSato2014}.

From this viewpoint, searching for symmetry-enforced constraints is practical.
For instance, transformation rules of Wilson-loops under group actions provide
additional global constraints on the allowed holonomies.
Such constraints reduce the amount of information that has to be supplied
independently, and hence provide a natural route to break the information
hierarchy~\cite{ShiozakiSciPost2026Equivariant,watanabe2026exactconjugationidentitymanybody}.

The other yes-go scenario, namely the case of quadratic Hamiltonians, also helps explain the success of prior works.
For instance, the pioneering density-functional treatment of polarization in noninteracting band insulators relies on a single-particle description \cite{KingSmithVanderbilt1993}.
More generally, if an interacting ground state is well approximated by a quasi-free state, approaches based on the one-body reduced density matrix (1RDM) may remain effective \cite{GiesbertzRuggenthaler2019PhysRep,RequistGross2018Polarization}.
This effectiveness reflects the absence or perturbative weakness of
interactions: the many-body problem reduces to independent single-particle
problems with controlled corrections.

In principle, an information hierarchy may also be expected in the scattering problem.
The scattering phase shift can be viewed as a global phase shift in the wavefunction accumulated under an evolution from $t=0$ to $t\to\infty$ \cite{PeskinSchroeder1995}.
In practice, however, scattering theory often relies on perturbative expansions around quasi-free asymptotic states, and additional symmetry constraints can further restrict the allowed phase shifts.
The practical success of scattering theory does not negate the presence of an information hierarchy.
Rather, it provides another perspective on why yes-go scenarios can arise in concrete settings.

The information hierarchy indicates a limitation of machine-learning techniques.
As we have shown, a finite set of constraints on $r$-point density correlators does not determine the many-body Berry phase.
This remains true even if one allows an arbitrary functional constraint.
Since a neural network is fundamentally a nonlinear functional of its inputs, it effectively imposes a constraint of the form
\[
f\bigl(\langle \hat n\rangle,\langle \hat n\hat n\rangle,\langle \hat n\hat n\hat n\rangle,\ldots\bigr)=\mathrm{const.}
\]
Accordingly, the many-body holonomy cannot be reproduced in general from such finite local correlators.
Nonlinearity does not help unless additional, genuinely nonlocal input is provided or the hypothesis class is restricted to a yes-go regime.
A simple way around this obstruction is to train on data originating from symmetric many-body systems, or to include global holonomies as input features together with density correlators~\cite{OrtizSouzaMartin1998ExchangeCorrelationHole}.
One promising way to obtain such global data would be to directly acquire data such as polarization through experiments~\cite{NakatsujiKiyoharaHigo2015Nature,WalterZhuGaechterMinguzziRoschinskiSandholzerViebahnEsslinger2023NatPhys}.
Indeed, many recent machine-learning studies aimed at predicting phase diagrams use, either explicitly or implicitly, nonlocal inputs or data drawn from symmetry-constrained settings \cite{CarrasquillaMelko2017NatPhys,vanNieuwenburg2017NatPhys,ZhangShen2018PRL,SunYiYang2018PRB}.
In such cases the problem is effectively restricted to a yes-go regime.

\section{Summary and outlook}
\label{sec:summary}
In this work, we studied to what extent the geometric information of an interacting many-body ground state can be inferred from a finite set of density correlations.
Using the Resta formula and viewing $\log z$ as a cumulant generating functional, we established a generic cumulant order-by-order information hierarchy.
In general, low-order information does not determine higher-order information.
That is, fixing density correlators (or cumulants) up to a finite order does not, in general, constrain higher-order ones, so the many-body holonomy is not uniquely determined from such low-order data.

In the thermodynamic limit, the information hierarchy is essentially local-to-global.
Finite-order local correlators constrain the local structure of the state, but
they do not determine the global structure in general.

We also identified two yes-go mechanisms that break this hierarchy.
First, for quasi-free (Gaussian) states, the hierarchy effectively closes and $z$ is determined by low-order data.
Second, symmetry constraints can reduce the infinite cumulant sum entering $\log z$ into finite information, which collapses the hierarchy.

From a practical viewpoint, our results identify what must be supplied, beyond finite-order density correlators, in order to access many-body holonomy.
In yes-go cases, symmetry constraints or (approximate) quasi-free structure compress the relevant information into a finite set of descriptors, so that $\gamma$ becomes accessible from reduced data.

Outside such regimes, no finite-order local information is sufficient, and one must incorporate genuinely nonlocal input or evaluate the global holonomy directly.
In this regard, methods that directly target many-body wave functions, such as DMRG~\cite{White1992,HauschildPollmann2018} and quantum Monte Carlo~\cite{Foulkes2001RMP}, are the representatives.
In the longer term, quantum computers may also provide a route to access such global holonomy data~\cite{NiedermeierQC2024,TamiyaVQEBP2021,Kattemolle2022PRB106_214429}, provided that the hardware and algorithms mature sufficiently.
From this perspective, data-driven approaches would ultimately require inputs that contain the many-body holonomy itself, or other genuinely nonlocal information that is equivalent to it.

\clearpage
\appendix
\appendix
\begin{widetext}
\section{$\theta$-independent bulk effective description and the gap non-closure}
\label{app:gap_and_seam}
We start from two $\theta$-threaded Hamiltonians,
$\hat H_{E}(\theta)$ and $\hat H_{O}(\theta)$, defined on chains with the same number of sites and satisfying
\begin{equation}
\hat H_{E}(\theta)\,|\Psi_{E}(\theta)\rangle
=
E_{E}(\theta)\,|\Psi_{E}(\theta)\rangle,
\qquad
\hat H_{O}(\theta)\,|\Psi_{O}(\theta)\rangle
=
E_{O}(\theta)\,|\Psi_{O}(\theta)\rangle.
\end{equation}
Here $E$ and $O$ label two distinct Hamiltonians. Accordingly, the Berry phases computed from these Hamiltonians are, in general, different, as seen in the main text. Concretely, one may think of two Hamiltonians whose ground states have even and odd bulk particle-number parity, respectively.

The gap along the $\theta$-cycle is then given by
\begin{equation}
\Delta(\theta)=E_{O}(\theta)-E_{E}(\theta).
\end{equation}
As discussed in the main text, $\theta$ is $2\pi$-periodic.

At $\theta=0$, we define the reference eigenstates and the bulk gap by
\begin{equation}
\hat H_{E}(0)\,|\Psi_E\rangle = 0,
\qquad
\hat H_{O}(0)\,|\Psi_O\rangle = \Delta\,|\Psi_O\rangle,
\end{equation}
so that $\Delta=\Delta(0)$ is the energy gap between the even and odd ground states at $\theta=0$.

The assumption in the main text is that the energy gap $\Delta$ is sufficiently large to separate the even and odd ground states.
Below we show that, under appropriate conditions, the gap does not close along the $\theta$-cycle.
Therefore, the Berry phase discussed in the main text is well defined, and our construction indeed provides a counterexample.

\subsection{Edge--bulk bipartition and Schmidt decomposition}
\label{app:schmidt_decomp_bulk}
We show that the $\theta$-dependence of the energy gap is sufficiently small, and that this conclusion is independent of the gauge fixing.

Adopting a twisted-boundary gauge, we choose the edge region to carry all explicit
$\theta$-dependence. Conversely, the bulk is chosen to be $\theta$-independent.
Symbolically, this bipartition is written as
\begin{equation}
\mathcal H=\mathcal H_{\rm edge}\otimes \mathcal H_{\rm bulk}.
\label{eq:Hilbert_split_app}
\end{equation}
Then, for each sector the ground state $|\Psi_s(\theta)\rangle$ is decomposed by the Schmidt decomposition as
\begin{equation}
|\Psi_s(\theta)\rangle
=
\sum^{\chi_s}_{\alpha= 1}\lambda^{(s)}_{\alpha}(\theta)\,
|e^{(s)}_{\alpha}(\theta)\rangle\otimes|b^{(s)}_{\alpha}\rangle,
\qquad
\lambda^{(s)}_{\alpha}(\theta)\ge 0.
\label{eq:schmidt_app}
\end{equation}
Here, $|e^{(s)}_{\alpha}(\theta)\rangle\in\mathcal H_{\rm edge}$ denotes the edge Schmidt state, and
$|b^{(s)}_{\alpha}\rangle\in\mathcal H_{\rm bulk}$ denotes its Schmidt partner.
The sector label $s\in\{E,O\}$ distinguishes the even/odd ground states.

The Schmidt sum over $\alpha$ is truncated at a bond dimension $\chi_s$.
In practice, it is chosen large enough to faithfully capture the entanglement relevant.
However, in this Appendix we do not rely on any finite-bond-dimension truncation and formally take the Schmidt
expansion without truncation (equivalently, $\chi_s\to\infty$ for each sector $s$).
In the following, we omit $\chi_s$ and treat the Schmidt expansion as untruncated.

The states satisfy the orthonormality relations
\begin{equation}
\langle e^{(s)}_{\alpha}(\theta)|e^{(s)}_{\beta}(\theta)\rangle=\delta_{\alpha\beta},
\qquad
\langle b^{(s)}_{\alpha}|b^{(s)}_{\beta}\rangle=\delta_{\alpha\beta}.
\label{eq:schmidt_ortho_app}
\end{equation}

Such a choice of edge and bulk states is possible because the seam position can be relocated by a gauge choice.
The bulk states and bulk operators can be isolated from the gauge-dependent edge states, leaving at most an exponentially suppressed $\theta$-dependence~\cite{WatanabeHA}.

\subsection{Bulk-space operator description}
\label{app:S_deltaS}
Now we introduce an effective bulk-space approach by defining effective operators acting on the bulk.
This approach is closely related in spirit to Ref.~\cite{Schmidt_ET}, although our construction is not identical.

Recall that the ground-state energy is given by the expectation value
\begin{equation}
E_s(\theta)
=
\langle\Psi_s(\theta)|\,\hat H_s(\theta)\,|\Psi_s(\theta)\rangle,
\qquad s\in\{E,O\}.
\label{eq:Es_expectation_app}
\end{equation}
Substituting the Schmidt decomposition \eqref{eq:schmidt_app} into \eqref{eq:Es_expectation_app}, we obtain
\begin{align}
E_s(\theta)
&=
\sum_{\alpha,\beta\ge 1}
\lambda^{(s)}_{\alpha}(\theta)\lambda^{(s)}_{\beta}(\theta)\,
\Bigl(
\langle e^{(s)}_{\alpha}(\theta)|\otimes
\langle b^{(s)}_{\alpha}|
\Bigr)\,
\hat H_s(\theta)\,
\Bigl(
|e^{(s)}_{\beta}(\theta)\rangle\otimes
|b^{(s)}_{\beta}\rangle
\Bigr)
\nonumber\\
&=
\sum_{\alpha,\beta\ge 1}
\lambda^{(s)}_{\alpha}(\theta)\lambda^{(s)}_{\beta}(\theta)\,
\langle b^{(s)}_{\alpha}|\,
\Bigl(
\langle e^{(s)}_{\alpha}(\theta)|\,
\hat H_s(\theta)\,
|e^{(s)}_{\beta}(\theta)\rangle
\Bigr)\,
|b^{(s)}_{\beta}\rangle.
\label{eq:Es_expand_app}
\end{align}

Equation \eqref{eq:Es_expand_app} motivates introducing the bulk-space operator
\begin{equation}
\hat S^{(s)}_{\alpha\beta}(\theta)
\equiv
\langle e^{(s)}_{\alpha}(\theta)|\,\hat H_s(\theta)\,|e^{(s)}_{\beta}(\theta)\rangle.
\label{eq:S_def_app}
\end{equation}
For each fixed $(s,\alpha,\beta)$, $\hat S^{(s)}_{\alpha\beta}(\theta)$ is an operator acting on
 the bulk-space $\mathcal H_{\rm bulk}$.
In terms of \eqref{eq:S_def_app}, Eq.~\eqref{eq:Es_expand_app} reads
\begin{equation}
E_s(\theta)
=
\sum_{\alpha,\beta\ge 1}
\lambda^{(s)}_{\alpha}(\theta)\lambda^{(s)}_{\beta}(\theta)\,
\langle b^{(s)}_{\alpha}|\,\hat S^{(s)}_{\alpha\beta}(\theta)\,|b^{(s)}_{\beta}\rangle.
\label{eq:Es_in_terms_of_S_app}
\end{equation}

We further separate the explicit $\theta$-dependence of the Hamiltonian by
\begin{equation}
\delta \hat H_s(\theta):=\hat H_s(\theta)-\hat H_s(0),
\label{eq:deltaH_def_app}
\end{equation}
and define the corresponding decomposition of $\hat S^{(s)}_{\alpha\beta}(\theta)$ as
\begin{equation}
\delta \hat S^{(s)}_{\alpha\beta}(\theta)
:=
\langle e^{(s)}_{\alpha}(\theta)|\,\delta\hat H_s(\theta)\,|e^{(s)}_{\beta}(\theta)\rangle,
\label{eq:deltaS_def_app}
\end{equation}
\begin{equation}
\hat S^{(s,0)}_{\alpha\beta}(\theta)
:=
\langle e^{(s)}_{\alpha}(\theta)|\,\hat H_s(0)\,|e^{(s)}_{\beta}(\theta)\rangle,
\label{eq:S0_def_app}
\end{equation}
so that
\begin{equation}
\hat S^{(s)}_{\alpha\beta}(\theta)
=
\hat S^{(s,0)}_{\alpha\beta}(\theta)
+
\delta \hat S^{(s)}_{\alpha\beta}(\theta).
\label{eq:S_split_app}
\end{equation}
Equation \eqref{eq:S_split_app} is an identity following from \eqref{eq:deltaH_def_app}.

The point is that, once one insists on a bulk-space operator description through \eqref{eq:S_def_app},
$\theta$-dependence necessarily enters the bulk effective model in two ways: (i) through the explicit
$\theta$ dependent part of the Hamiltonian $\delta\hat H_s(\theta)$, which contributes via \eqref{eq:deltaS_def_app},
and (ii) through the $\theta$-dependence of the edge Schmidt states $|e^{(s)}_{\alpha}(\theta)\rangle$ used
to define \eqref{eq:S_def_app}, reflecting the cut-induced (virtual) gluing across the gauge-dependent
edge--bulk decomposition.

The latter implies that although $\hat H_s(0)$ itself has no $\theta$-dependence, the edge degrees of freedom are coupled to the flux.
This coupling enters to the bulk-space operator $\hat S^{(s,0)}_{\alpha\beta}(\theta)$ as a correction associated with the artificial decomposition.
Similar cut-induced contributions, where an effective subspace inherits nontrivial contributions from its Schmidt-complement space, have been discussed in the contexts of matrix-product-state constructions and many-body entanglement, e.g., Refs.~\cite{Schmidt_ET,Sommer_Wen_Vishwanath_HBC_I_2025PRL,Zhang_OperatorSchmidt_PRL_2024}.

This idea of virtual current mechanism has a close analogue in effective descriptions in nuclear physics, where integrating out
degrees of freedom of the exchanged charged-mesons induces ``exchange-current'' contributions~\cite{KOICHI1989564,WK2018EC}.
For example, in an effective two-neutron description written solely in terms of an $NN$ potential, one may
naively expect that an external electromagnetic field has no effect to the system.
However, the $NN$ potential itself encodes pion-exchange processes that have been integrated out, and therefore induces currents in the effective theory (often discussed as exchange currents).

\subsection{Bulk gap non closure}
\label{app:gap_nonclosure}
We now fix a common gauge for the even/odd states by aligning the position of the twisted boundary condition (tBC).
The tBC position is gauge-dependent and can be relocated by a gauge transformation.
Concretely, the twist implemented at a link $\ell$ can be shifted to another link $\ell'$ by
\begin{equation}
\hat U_{\ell\to \ell'}(\theta)
:=
\begin{cases}
\displaystyle
\exp\!\left(
-i\theta\sum_{j=\ell+1}^{\ell'} \hat n_j
\right),
& \ell<\ell',
\\[1.2em]
\displaystyle
\exp\!\left(
+i\theta\sum_{j=\ell'+1}^{\ell} \hat n_j
\right),
& \ell>\ell' .
\end{cases}
\label{eq:U_shift_twist_app}
\end{equation}
For example, consider a gauge in which the twist is localized on the bond $(\ell,\ell+1)$,
\begin{equation}
\hat h^{(\ell)}_{\rm tBC}(\theta):=t\,e^{-i\theta}\,\hat c^\dagger_{\ell}\hat c_{\ell+1}+{\rm h.c.}
\label{eq:tBC_only}
\end{equation}
and all other hopping terms carry no phase.
The unitary tBC shift acts on fermion operators as follows.
For $\ell<\ell'$ and $\ell<j<\ell'$,
\[
\hat U_{\ell\to \ell'}(\theta)\,
\hat c_j\,
\hat U^{-1}_{\ell\to \ell'}(\theta)
=
e^{+i\theta}\hat c_j,
\qquad
\hat U_{\ell\to \ell'}(\theta)\,
\hat c_j^\dagger\,
\hat U^{-1}_{\ell\to \ell'}(\theta)
=
e^{-i\theta}\hat c_j^\dagger.
\]
For $\ell>\ell'$ and $\ell'<j<\ell$,
\[
\hat U_{\ell\to \ell'}(\theta)\,
\hat c_j\,
\hat U^{-1}_{\ell\to \ell'}(\theta)
=
e^{-i\theta}\hat c_j,
\qquad
\hat U_{\ell\to \ell'}(\theta)\,
\hat c_j^\dagger\,
\hat U^{-1}_{\ell\to \ell'}(\theta)
=
e^{+i\theta}\hat c_j^\dagger.
\]
Hence the hopping on any internal bond $(j,j+1)$ with $\ell<j<\ell'$ is unchanged, since the phases cancel:
$\hat c^\dagger_j\hat c_{j+1}\mapsto (e^{\pm i\theta}\hat c^\dagger_j)(e^{\mp i\theta}\hat c_{j+1})=\hat c^\dagger_j\hat c_{j+1}$, where the choice of signs follows from the two cases $\ell<\ell'$ and $\ell>\ell'$ in Eq.~\eqref{eq:U_shift_twist_app}.
Thus, the phase factors cancel for both $\ell<\ell'$ and $\ell>\ell'$.
On the other hand, the boundary bonds transform as
\[
\hat c^\dagger_{\ell}\hat c_{\ell+1}\mapsto \hat c^\dagger_{\ell}(e^{+i\theta}\hat c_{\ell+1})
=e^{+i\theta}\hat c^\dagger_{\ell}\hat c_{\ell+1},
\]
for $\ell<\ell'$, or,
\[
\hat c^\dagger_{\ell}\hat c_{\ell+1}\mapsto (e^{+i\theta}\hat c^\dagger_{\ell})\hat c_{\ell+1}
=e^{+i\theta}\hat c^\dagger_{\ell}\hat c_{\ell+1},
\]
for $\ell>\ell'$.
Either way, acting with the unitary $\hat U_{\ell\to\ell'}(\theta)$ on the Hamiltonian removes the twist from the bond $(\ell,\ell+1)$ given by \Eq{eq:tBC_only}.
Similar calculations show that the phase is relocated to the bond $(\ell',\ell'+1)$.
Thus, the unitary shifts the tBC from $\ell$ to $\ell'$.
We therefore can choose a gauge in which the tBC position is common to the even and odd ground states.

The neighborhood of the aligned tBC position is nothing but the edge region, the bulk is the complement region sufficiently far from the tBC.
With this tBC position, we assume that the operator
$\hat S^{(s,0)}_{\alpha\beta}(\theta)$ introduced in Eq.~\eqref{eq:S_def_app} can be chosen to be sector-independent, namely
\begin{equation}
\hat S^{(s,0)}_{\alpha\beta}(\theta)=\hat S_{\alpha\beta}(\theta).
\label{eq:S_sector_indep_assump_app}
\end{equation}
Moreover, the $\theta$-dependence of $\hat S_{\alpha\beta}(\theta)$ is due to the virtual contribution from the edge state $|e_\beta(\theta)\rangle$.
Any remaining $\theta$-dependence in the bulk effective description originates solely from the virtual current.
We assume that this contribution is exponentially suppressed in the system size, following Ref.~\cite{WatanabeHA}.
Hence, we can rewrite
\begin{equation}
\hat S_{\alpha\beta}(\theta) \simeq \hat S_{\alpha\beta}+\mathcal O(e^{-L/\xi}).
\label{eq:S_sector_indep_assump_app_exp}
\end{equation}
Here, $\xi$ is the correlation length, which is typically smaller than the system size $L$ for a one-dimensional system.
On the other hand, the gauge-dependent edge contribution (contribution from $\delta \hat S_{\alpha\beta}(\theta)$) is common to the even and odd states and therefore cancels in their difference.

Since we adopt the same aligned tBC position for the even and odd states, it is not a severe assumption to also assume that the Schmidt coefficients are common.
By isolating the Schmidt cut from the tBC, the remaining $\theta$-dependence is suppressed as well.
The resulting expression is
\begin{equation}
\lambda^{(s)}_{\alpha}(\theta)\simeq\lambda_{\alpha}+\mathcal O(e^{-L/\xi}).
\label{eq:lambda_sector_indep_assump_app}
\end{equation}

We now evaluate the even--odd energy difference using Eq.~\eqref{eq:Es_in_terms_of_S_app}.
Substituting equations Eqs. (\ref{eq:S_sector_indep_assump_app_exp}) and (\ref{eq:lambda_sector_indep_assump_app}) we obtain
\begin{equation}
E_O(\theta)-E_E(\theta)
\simeq
\sum_{\alpha,\beta\ge 1}
\lambda_{\alpha}\lambda_{\beta}\left\{ \,
\langle b^{(O)}_{\alpha}|\hat S_{\alpha\beta}|b^{(O)}_{\beta}\rangle
-
\langle b^{(E)}_{\alpha}|\hat S_{\alpha\beta}|b^{(E)}_{\beta}\rangle
\right\} + \mathcal O(e^{-L/\xi}).
\label{eq:EOEE_S_leading_app}
\end{equation}
This implies that the $\theta$-dependence on the even and odd energy difference is at most
exponentially small in $L$ once the tBC position is aligned, consistent with Haruki Watanabe's observation.

Therefore, the even--odd splitting is mainly dominated by its value at $\theta=0$,
\begin{equation}
E_O(\theta)-E_E(\theta)=\bigl[E_O(0)-E_E(0)\bigr]+O(e^{-L/\xi}),
\label{eq:EOEE_theta_small_app}
\end{equation}
and since $\Delta=E_O(0)-E_E(0)>0$ is assumed to be sufficiently large in the beginning, the gap cannot close
along the $\theta$-cycle.

\subsection{Remark: Hellmann--Feynman Theorem}
\label{app:HF_remark}
Let us leave a comment on the Hellmann--Feynman theorem.
Naively, the Hellmann--Feynman theorem seems useful for evaluating the excitation gap.
A direct application gives
\[
\partial_\theta\!\left(E_O(\theta)-E_E(\theta)\right)
=
\langle\Psi_O(\theta)|\,\partial_\theta \hat H_O(\theta)\,|\Psi_O(\theta)\rangle
-
\langle\Psi_E(\theta)|\,\partial_\theta \hat H_E(\theta)\,|\Psi_E(\theta)\rangle.
\]
If one could identify $\hat H_O(\theta)-\hat H_E(\theta)$ as a purely bulk energy difference and, at the same time,
compare the expectation values in the same gauge, this would suggest a $\theta$-independent gap.
However, the $\theta$-dependence of $|\Psi_s(\theta)\rangle$ is not unique under tBC relocation, and the
bulk/edge separation becomes meaningful only after fixing an explicit gauge-dependent decomposition,
as discussed above.

The naive Hellmann--Feynman argument assumes that the relevant state can be
compared as a normalized eigenstate in a fixed Hilbert-space representation.
In the present bulk-space description, however, the bulk states are kept
$\theta$-independent, while their Schmidt counterparts, namely the edge Schmidt
states, carry the $\theta$-dependence. Hence, the effective bulk expectation
value contains additional $\theta$-dependent weights and matrix elements, and
the normalization argument used in the standard derivation of the
Hellmann--Feynman theorem does not directly apply to the bulk-effective
description.
Hence, this naive argument is not sufficient by itself to establish the gap non-closure.
\end{widetext}

\begin{widetext}
\section{Alternative proof based on the occupation-basis factorization}
\label{app:factorization}
We prove the information hierarchy by directly constructing the occupation-basis counterexample.
As discussed in the main text, the ground state in the occupation basis is given by
\begin{equation}
|\Psi\rangle=\sum_{n_1,\dots,n_L\in\{0,1\}} a_{n_1\cdots n_L}\,|n_1\cdots n_L\rangle.
\end{equation}

We apply the Resta formula to two distinct ground states $|\Psi\rangle$ and $|\Psi'\rangle$, namely
\begin{equation}
z=\langle \Psi|e^{i\alpha_L \hat X}|\Psi\rangle,
\qquad
z'=\langle \Psi'|e^{i\alpha_L \hat X}|\Psi'\rangle.
\end{equation}

Since $\hat n_j$ is diagonal in the occupation basis and satisfies $\hat n_j^2=\hat n_j$, the operator
\[
\hat X=\sum_{j=1}^{L} j\,\hat n_j
\]
is also diagonal, and its moments are directly expressed in terms of density correlators.
Moreover, keeping $\hat n_j^2 = \hat n_j$ in mind, the exponential appearing in the Resta expression simplifies as
\begin{equation}
e^{i\alpha_L j\hat n_j}=1+\bigl(e^{i\alpha_L j}-1\bigr)\hat n_j.
\end{equation}
Hence one can factorize the exponential and expand it site by site,
which yields
\begin{equation}
z
=1+\sum_{p=1}^{L}\ \sum_{1\le j_1<\cdots<j_p\le L}
\Bigg[\prod_{m=1}^{p}\bigl(e^{i\alpha_L j_m}-1\bigr)\Bigg]
A_{j_1\cdots j_p},
\end{equation}
where $\langle \hat n_{j_1}\cdots \hat n_{j_p}\rangle=A_{j_1\cdots j_p}$.
An identical expression for $z'$ with $A_{j_1\cdots j_p}$ replaced by $A'_{j_1\cdots j_p}$.

Suppose that all $p$-point correlators with $p\le r$ are identical for the states $|\Psi\rangle$ and $|\Psi'\rangle$.
Subtracting the two expansions, we obtain
\begin{equation}
z-z'
=\sum_{p=r+1}^{L}\ \sum_{1\le j_1<\cdots<j_p\le L}
\Bigg[\prod_{m=1}^{p}\bigl(e^{i\alpha_L j_m}-1\bigr)\Bigg]
\Bigl(A_{j_1\cdots j_p}-A'_{j_1\cdots j_p}\Bigr).
\end{equation}
Thus, the leading contribution is controlled by the $(r+1)$-point difference,
while there is no reason for higher-point correlators to coincide in general, in close analogy with the $L=6$ example discussed above.
For example, even the connected density $r+1$-point function in $z$ and $z'$ can differ, namely, $A_{1\cdots r+1} \neq A'_{1\cdots r+1}$, and the Berry phase may differ as well.
\end{widetext}

\bibliography{refs}
\end{document}